\RequirePackage{fix-cm}
\documentclass[twocolumn,epjc3]{svjour3}  
\smartqed  
\RequirePackage{graphicx}
\RequirePackage{latexsym}
\RequirePackage[numbers,sort&compress]{natbib}
\RequirePackage[colorlinks,citecolor=blue,urlcolor=blue,linkcolor=blue]{hyperref}
\RequirePackage{amssymb}
\RequirePackage{amsmath}
\RequirePackage{braket}
\RequirePackage{hyperref}
\RequirePackage{bigints}
\RequirePackage{csquotes}
\RequirePackage[dvipsnames]{xcolor}
\journalname{Eur. Phys. J. A}

\begin{document}

\title{Gradient expansion technique for inhomogeneous, magnetized quark matter
}


\author{Filippo Anzuini\thanksref{addr1, e1}
        \and
        Andrew Melatos\thanksref{addr1,addr2} 
}

\thankstext{e1}{e-mail: fanzuini@student.unimelb.edu.au}


\institute{School of Physics, University of Melbourne, Parkville, Victoria 3010, Australia \label{addr1}
           \and
           Australian Research Council Centre of Excellence for Gravitational Wave Discovery (OzGrav), University of Melbourne, \\
 \  Parkville, Victoria 3010, Australia \label{addr2}
}

\date{Received: date / Accepted: date}

\maketitle

\begin{abstract}
A quark-magnetic Ginzburg-Landau (qHGL) gradient expansion of the free energy of two-flavor inhomogeneous quark matter in a magnetic field $H$ is derived analytically. It can be applied away from the Lifshitz point, generalizing standard Ginzburg-Landau techniques. The thermodynamic potential is written as a sum of the thermal contribution, the non-thermal lowest Landau level contribution, and the non-thermal qHGL functional, which handles any arbitrary position-dependent periodic modulation of the chiral condensate as an input. The qHGL approximation has two main practical features: (1) it is fast to compute; (2) it applies to non-plane-wave modulations such as solitons even when the amplitude of the condensate and its gradients are large (unlike standard Ginzburg-Landau techniques). It agrees with the output of numerical techniques based on standard regularization schemes and reduces to known results at zero temperature ($T = 0$) in benchmark studies. It is found that the region of the $\mu$-$T$ plane (where $\mu$ is the chemical potential) occupied by the inhomogeneous phase expands, as $H$ increases and $T$ decreases.

\end{abstract}

\section{Introduction}
\label{Introduction}
Quarks acquire an effective mass due to spontaneous chiral symmetry breaking \cite{Klevansky_1992, Buballa_2005, Nickel_2009, Abuki_2012, Buballa_2015}, which occurs in analogy to the BCS theory of superconductivity \cite{Bardeen_1957}. In vacuo, quark-antiquark pairs with opposite chirality condense, while in dense matter the chiral condensate forms due to the coupling of particles and holes \cite{Buballa_2005, Kojo_2010, Buballa_2015}. The condensate is homogeneous when the Cooper pairs form at the Fermi surface with a total null momentum. Inhomogeneity arises when the chiral condensate acquires a net momentum. For example in dense matter the pairing of particles and holes with equal and opposite momenta is energetically disfavored, and is superseded by the pairing of particles and holes with similar, aligned momenta \cite{Kojo_2010, Buballa_2015}. Inhomogeneous phases are commonly found in effective-model calculations \cite{Kutschera_1990, Nickel_2009,Fukushima_2011, Abuki_2012, Buballa_2015}.

Several studies have focused on the influence of external magnetic fields on quark matter \cite{Fraga_2008, Frolov_2010, Ferrer_2010, Gatto_2010, Andersen_2014, Miransky_2015, Tatsumi_2015, Menezes_2016, Abuki_2018}. Extensive research has been devoted to the effect of strong magnetic fields ($H \gtrsim 10^{16} \  \textrm{G}$) on inhomogeneous condensates, on the phase diagram of quantum chromodynamics (QCD) \cite{Frolov_2010, Fukushima_2011, Carignano_2015, Tatsumi_2014, Tatsumi_2015, Buballa_2016, Abuki_2018}, on neutrino emission from magnetized quark matter \cite{Xue_Wen_2007} and on magnetic catalysis \cite{Klevansky_1989, Klimenko_2005, Fukushima_2013, Andersen_2014, Miransky_2015}. The latter refers to the increase of the quark effective mass due to strong, external magnetic fields.

In this work we study the phase diagram of magnetized, hot quark matter for magnetic fields $H \lesssim 10^{18}$ G and temperatures $T \lesssim 30$ MeV that may occur in neutron star mergers \cite{Radice_2018,Perego_2019, Endrizzi_2020}) or that can be probed in laboratory experiments \cite{Jacobs_2005,Adams_2005}. We derive an analytic gradient expansion of the free energy of two-flavor quark matter in moderate magnetic fields ($H \lesssim 10^{18}$ G), extending the results found in \cite{Carignano_2018} for unmagnetized, inhomogeneous chiral condensates. The gradient expansion is referenced henceforth by the acronym qHGL, standing for quark-magnetic ($H$)-Ginzburg- Landau. Unlike the standard Ginzburg-Landau expansion \cite{Nickel_2009,Abuki_2012, Buballa_2015}, the qHGL approximation can be applied away from the Lifshitz point, i.e. in regions of the phase diagram where the amplitude of the condensate and its gradients are large. It is fast to compute and can be applied to plane-wave and non-plane-wave modulations of the condensate (such as solitonic modulations for example). It accurately describes first- and second-order phase transitions of the homogeneous chiral condensate to the inhomogeneous phase and of second-order phase transitions to the chirally restored phase. It also yields the amplitude and wave vector of the chiral condensate required to calculate neutrino emissivities of inhomogeneous \cite{Tatsumi_2014} and magnetized quark matter \cite{Xue_Wen_2007} for example. For magnetic fields of the order of $H \sim 10^{17}$ G (which are plausible in the core regions of some neutron stars with surface magnetic fields near the upper end of the range $10^{13} \ \textrm{G} \lesssim H \lesssim 10^{15} \ \textrm{G}$ \cite{Lyne_2006,Menezes_2016, Potekhin_2020}), many Landau levels are occupied, and the qHGL functional determines what condensate shape is thermodynamically favored, enriching the phase diagram for quark matter in the presence of external magnetic fields and simplifying previous numerical studies \cite{Nickel_2009, Buballa_2015, Nishiyama_2015, Cao_2016}. 

For $T > 0$ the thermodynamic potential acquires an additional thermal contribution. The qHGL expansion approximates the non-thermal contribution of the higher Landau levels. Previous works \cite{Tatsumi_2015, Abuki_2018} study thermal and magnetic effects close to the Lifshitz point, i.e. for high temperatures ($T \approx T_c$, where $T_c$ is the critical temperature at which the condensate melts), and often consider ultra-strong magnetic fields ($\sqrt{eH} \gtrsim 100$ MeV). We study the competition of magnetic and thermal effects in the regime $T \sim \sqrt{eH}$ (with $10 \ \textrm{MeV} \lesssim T \lesssim 30$ MeV), which may be reached in astrophysical environments or laboratory experiments \cite{Radice_2018,Perego_2019, Endrizzi_2020}. The inhomogeneous phase shifts to lower chemical potentials, and the density range occupied by the nonuniform chiral condensate shrinks, as $T$ increases and $H$ decreases.

The paper is structured as follows. Section \ref{sec:NJL_model} introduces the effective model adopted in this work and describes how the quark energy levels are modified in the presence of an external magnetic field. The  qHGL functional is derived in Section \ref{sec:qHGL_expansion}. Section \ref{sec:QCD_phase_diagram} studies the phase diagram for quark matter at high density and at $T = 0$ for a plane-wave and a solitonic shape modulation of the quark condensate using the qHGL functional, comparing unmagnetized and magnetized systems. In Section \ref{sec:QCD_phase_diagram_T} we consider a plane-wave modulation and show how the chiral condensate changes for $T>0$ and $H > 0$.

\section{Nambu-Jona Lasinio model}
\label{sec:NJL_model}
The Nambu-Jona Lasinio (NJL) model \cite{Nambu_1961a, Nambu_1961b}, firstly proposed to describe the dynamical generation of the nucleon mass in analogy to the BCS theory of superconductivity \cite{Bardeen_1957}, has been widely applied to quark matter \cite{Klevansky_1992, Buballa_2005, Nickel_2009, Carignano_2015, Buballa_2015, Buballa_2016}. The NJL model preserves the symmetries of QCD and predicts the generation of the quark mass via the mechanism of chiral symmetry breaking, which occurs due to the coupling of quarks and antiquarks with opposite chiralities (in vacuo). The interaction among quarks and antiquarks is described in terms of contact coupling, which is nonrenormalizable and must be regularized to remove unphysical ultraviolet divergencies \cite{Klevansky_1992}.

\subsection{Lagrangian density}

The Lagrangian density of the NJL model in the presence of an external magnetic field is

\begin{eqnarray}
\mathcal{L} = \bar \psi\Big(i\gamma^{\mu}D_{\mu} - m_q\Big)\psi + G_{S}\Big[(\bar \psi \psi)^2 + (\bar\psi i \gamma^5 \vec{\tau}  \psi)^2\Big] , \
\label{eq:NJL_lagrangian}
\end{eqnarray}
where $\psi$ is the quark doublet field in flavor space with three colors, $m_q = \textrm{diag}(m_u, m_d)$ is the quark mass matrix ($m_u$ and $m_d$ are the bare masses of the up and down quarks respectively), and $\gamma^{\mu}$ and $\gamma^5$ are the Dirac matrices. The covariant derivative is $D_{\mu} = \partial_{\mu} + i Q A_{\mu}$, where $Q = \textrm{diag}(2e/3, -e/3)$ is the electric charge matrix in flavor space, $e$ is the electronic charge and $A_{\mu}$ is the external electromagnetic four-potential in the Landau gauge, $A_{\mu} = \big(0,0,H x, 0\big)$\footnote{In the following we quote the values of $H$ in Gaussian units.}. The coupling constant $G_S$ regulates the strength of the scalar $(\bar \psi \psi)$ and pseudoscalar $(\bar\psi i \gamma^5 \tau^a  \psi)$ interaction terms, with $\tau^a$ denoting the isospin Pauli matrices in flavor space. In the following, we consider the chiral limit in which the bare quark mass vanishes and consider the case of symmetric quark matter, i.e. quark matter with equal concentrations of up and down quarks. Possible extensions to asymmetric quark matter in $\beta$-equilibrium, e.g. in neutron stars, are discussed in \ref{sec:Applications}.

We work in the mean-field approximation, in which the products of the quark field are replaced by $\bar\psi \psi = \braket{\bar{\psi}\psi} + \delta(\bar{\psi}\psi)$ (and similarly for the pseudoscalar term). The first term denotes the expectation value of the scalar condensate (which in general can depend on position \cite{Nakano_2005, Nickel_2009, Carignano_2015, Buballa_2015}); the second denotes fluctuations. Applying the mean-field approximation to Eq. (\ref{eq:NJL_lagrangian}), the quarks acquire an effective mass \cite{Nickel_2009, Buballa_2015}

\begin{eqnarray}
M(\bold{x}) = - 2 G_S \Big[\braket{\bar \psi \psi}(\bold{x}) + \braket{\bar\psi i \gamma^5 \tau^3  \psi}(\bold{x})\Big] \ ,
\label{eq:eff_mass}
\end{eqnarray}
where we assume that the condensates $\braket{\bar\psi i \gamma^5 \tau^1  \psi}$ and $\braket{\bar\psi i \gamma^5 \tau^2 \psi}$ vanish \cite{Klevansky_1992, Frolov_2010, Buballa_2015}. Keeping only $\braket{\bar\psi i \gamma^5 \tau^3 \psi}$ is equivalent to keeping only the exchange of neutral pions $\pi^0$ among quarks. We restrict to the case of periodic modulations of the chiral condensate \cite{Buballa_2015}.

\subsection{Magnetic field}
In the presence of an external magnetic field, the quark momenta perpendicular to the field direction are quantized \cite{Landau_1981}. Assuming that the effective mass is modulated as a plane wave (chiral density wave, CDW \cite{Kutschera_1990, Abuki_2012,Buballa_2015})

\begin{eqnarray}
M(z) = \Delta e^{2 i K z} \ ,
\label{eq:CDW}
\end{eqnarray}
with amplitude $\Delta$ and wave vector $2 K$, the energy levels for quarks of flavor $f$ are given by \cite{Frolov_2010, Carignano_2015, Buballa_2016}

\begin{equation}
E^f_{n,\zeta} = 
\begin{cases} 
\epsilon \sqrt{\Delta^2 + p_z^2} + K \qquad \qquad \qquad \qquad \qquad  n = 0 \\ 
 \\ 
\epsilon \sqrt{\big( \zeta \sqrt{\Delta^2 + p_z^2} + K \big)^2 + 2 |e_f H|n} \qquad \ n > 0 \\ 
\end{cases}
\label{eq:energy_levels}
\end{equation}
where $p_z$ is the quark momentum along the magnetic field direction, $p_{\perp} = \sqrt{2 |e_f H| n}$ is the quantized momentum perpendicular to the direction of the magnetic field \cite{Landau_1981}, $e_f$ is the charge of the quark with flavor $f$, $n$ enumerates the Landau level and we have $\epsilon, \zeta  = \pm$ (where $\zeta$ is the index for the quark spin). The energy of the higher Landau levels ($n > 0$) is symmetric, while the lowest Landau level (LLL, $n = 0$) is not. The spectral asymmetry is related to the axial anomaly \cite{Tatsumi_2015, Nishiyama_2015, Abuki_2018} and introduces an anomalous term in the quark free energy \cite{Frolov_2010, Abuki_2018, Buballa_2016}, as discussed in Section \ref{sec:Benchmark_Mag}.

\section{qHGL expansion}
\label{sec:qHGL_expansion}
The thermodynamically favored state for inhomogeneous quark matter is determined by finding the eigenvalues of the Hamiltonian operator and minimizing the corresponding free energy density \cite{Buballa_2015}. The numerical calculation is computationally demanding even for simple periodic modulations of the condensate. In Section \ref{sec:IGL} we briefly describe the improved Ginzburg-Landau (IGL) expansion \cite{Carignano_2018} for unmagnetized, inhomogeneous quark matter at zero temperature ($T = 0$), an analytic approximation of the quark free energy that simplifies the numerical study of inhomogeneous phases. In Section \ref{sec:qHGL} we extend the results found in \cite{Carignano_2018} to magnetized, inhomogeneous quark matter at $T = 0$. The case of nonzero temperatures is studied in Section \ref{sec:qHGL_thermal}.

\subsection{Unmagnetized quark matter at $T = 0$}
\label{sec:IGL}
The standard GL expansion of the quark thermodynamic potential \citep{Nickel_2009, Abuki_2012, Carignano_2018} is given by the sum of the thermodynamic potential $\mathcal{F}_{0}$ (which denotes the free energy density of uncondensed quarks) plus gradient terms of the quark condensate \cite{Nickel_2009, Abuki_2012, Carignano_2018}. The GL expansion can be applied only close to the Lifshitz point, i.e. where both the amplitude of the periodic condensate and its wave vector are small with respect to the chemical potential $\mu$. The IGL approximation (given by Eq. (2) in \cite{Carignano_2018}) improves the standard GL approximation (for vanishing temperatures) in two respects. First, the $\mathcal{F}_0$ term is replaced in the IGL approximation by the free energy of homogeneous quark matter $\mathcal{F}_{\hom} = \mathcal{F}_{\hom}(\overline{M^2(z)}, \mu)$ \cite{Carignano_2018}, which is a function of the moving average

\begin{eqnarray}
\overline{M^2(z)} = \frac{1}{\lambda}\int^{z+\frac{\lambda}{2}}_{z-\frac{\lambda}{2}}  dz^{\prime} M^2(z^{\prime}) \ .
\label{eq:moving_average}
\end{eqnarray}
In Eq. \eqref{eq:moving_average}, $\lambda$ is a typical wavelength for the quark condensate oscillations, which we fix to $\lambda = \mu^{-1}$ as in \citep{Carignano_2018}.
In the density region where the ground state is homogeneous, $\mathcal{F}_{\hom}(\overline{M^2(z)}, \mu)$ gives the correct free energy for a uniform condensate. On the other hand, it
reproduces the long-wavelength behavior of the condensate typical, for example, of second-order phase transitions (see \cite{Carignano_2018} for details). Second, the IGL functional includes the gradients in the standard GL expansion and contains additionally higher order gradients with respect to the standard GL expansion, which are
calculated from the $\mathcal{F}_2$ term in the following Taylor expansion for $\Delta \approx 0$ \cite{Carignano_2018}

\begin{eqnarray}
\mathcal{F} =  \ && \mathcal{F}_{0} + \mathcal{F} _{2}(K)\Delta^2 + \mathcal{F} _{4}(K)\Delta^4 + ...  \\ 
 =  \ && \mathcal{F}_{0} + \frac{\partial\mathcal{F}}{\partial(\Delta^2)}\Big|_{\Delta = 0} \Delta^2 + \frac{1}{2}\frac{\partial^2\mathcal{F}}{\partial(\Delta^2)^2}\Big|_{\Delta = 0}\Delta^4 \nonumber \\
&& + ... \ ,
\label{eqn:Kappaexp}
\end{eqnarray}
where the index of $\mathcal{F}_j$ denotes the associated power of $\Delta$.  The terms calculated from $\mathcal{F}_2$ are proportional to gradients of the form $|\nabla^j M|^2$, which dominate near second-order transitions, where $M$ is small, but $|\nabla^j M|^2$ is not ($\Delta \ll K \lesssim \mu$). Hence, contrarily to the standard GL expansion, the IGL approximation is valid at any $\mu$ in the phase diagram of quarks at zero temperature. The IGL functional, like the standard GL approximation, depends on $M$ and the gradients of $M$ without assuming the functional form in Eq. \eqref{eq:CDW} (cf. Eq.(2) in \cite{Carignano_2018}), and can be applied to any periodic modulation of the quark condensate \cite{Carignano_2018}.

\subsection{Magnetized quark matter at $T = 0$}
\label{sec:qHGL}
In this section we include an external magnetic field and derive an analytic approximation for the free energy associated with the higher Landau levels.
The total free energy density of the quarks can be written as
\begin{eqnarray}
\mathcal{F} = \mathcal{F}_{\textrm{LLL}} + \mathcal{F}_{\textrm{qHGL}} \ ,
\label{eq:F_MAG}
\end{eqnarray}
where the first term denotes the LLL contribution (which is proportional to $eH$ and which we calculate similarly to \cite{Frolov_2010}), and $\mathcal{F}_{\rm{qHGL}}$ is given by
\begin{eqnarray}
\mathcal{F}_{\rm{qHGL}} = \mathcal{F}^{\rm{vac}}_{\rm{ HLL}} + \mathcal{F}^{\rm{med}}_{\rm{ HLL}} + \mathcal{F}_{\rm{cond}} \ ,
\label{eq:F_HLL}
\end{eqnarray}
where the quantities on the right of Eq. \eqref{eq:F_HLL} denote the vacuum and medium free energy density associated with the higher Landau levels plus the condensation free energy density respectively \cite{Carignano_2015, Buballa_2015}. The latter is given by \cite{Buballa_2005, Buballa_2015}

\begin{eqnarray}
\mathcal{F}_{\rm{cond}} = \frac{|\Delta|^2}{4 G_S} \ .
\end{eqnarray}
The vacuum term diverges for high values of the quark momentum, and a regularization prescription is required. We adopt the Pauli-Villars scheme \citep{Klevansky_1992, Carignano_2018} to regularize $\mathcal{F}^{\rm{vac}}_{\rm{ HLL}}$. The Pauli-Villars scheme introduces fictitious, heavy particles of mass $\approx \Lambda$ \cite{Klevansky_1992}. It is implemented by replacing the quark energy levels in Eq. \eqref{eq:energy_levels} with 

\begin{equation}
    E^f_{n,\zeta}\mapsto E^f_{\textrm{PV}} = \sum_{k = 0}^3 c_k\sqrt{(E^f_{n,\zeta})^2+k\Lambda^2} \ ,
\end{equation}
with $c_0 = 1$, $c_1 = -3$, $c_2 = 3$, $c_3 = -1$ \citep{Klevansky_1992, Carignano_2018}. The regularized $\mathcal{F}^{\rm{vac}}_{\rm{ HLL}}$ reads \cite{Frolov_2010, Carignano_2015}

\begin{eqnarray}
\mathcal{F}^{\rm{vac}}_{\rm{ HLL}} = && -N_c \sum_{f} \frac{|e_f H|}{8 \pi^2} \bigintssss d p_{z}\!\!\! \sum_{n>0, \zeta, \epsilon}\!\!\!E^f_{\textrm{PV}} \ , 
\end{eqnarray}
where $N_c$ is the number of quark colors. For the medium contribution (which is not divergent) one has \cite{Frolov_2010, Carignano_2015}

\begin{eqnarray}
\mathcal{F}^{\rm{med}}_{\rm{ HLL}} = && -N_c \sum_{f} \frac{|e_f H|}{4 \pi^2} \times \nonumber \\
&& \bigintssss d p_{z}\!\!\! \sum_{n>0, \zeta}\!\! \Big(\mu - E^f_{n, \zeta}\Big)\Theta\Big(\mu - E^f_{n, \zeta}\Big)\Big{|}_{\epsilon = 1} \ , \nonumber 
\\
\label{eq:F_med}
\end{eqnarray}
where $\Theta$ denotes the Heaviside function.
We focus in the following on $\mathcal{F}^{\rm{med}}_{\rm{ HLL}}$ (the same procedure applies to $\mathcal{F}^{\rm{vac}}_{\rm{ HLL}}$). Expressing Eq. \eqref{eq:F_med} as a Taylor series in analogy to Eq.(\ref{eqn:Kappaexp}), we get (for details, cf. Eq. (10) in \cite{Carignano_2018})

\begin{eqnarray}
\frac{\partial \mathcal{F}^{\rm{med}}_{\rm{HLL}}}{\partial (\Delta^2)}\Big{|}_{\Delta = 0} = && N_c \sum_{f} \frac{|e_f H|}{4 \pi^2} \bigintssss d p_{z}\!\!\! \sum_{n>0, \zeta} \frac{1}{2 E^f_{n, \zeta}} \nonumber \\
&& \times \bigg(1 + \zeta \frac{K}{\sqrt{p_z^2}}\bigg)\Theta\Big(\mu - E^f_{n, \zeta}\Big)\Big{|}_{\epsilon = 1} \ .
\end{eqnarray}

If the magnetic field $H$ is small ($\sqrt{eH} \ll \mu$), many Landau levels are populated, and it is possible to approximate the sum over $n$ with an integration\footnote{In the literature (see for example \cite{Frolov_2010}) the continuum approximation is adopted for $\sqrt{eH} \ll \Lambda$.} \cite{Frolov_2010} starting from $p_{\perp} = \sqrt{2 |e_f H|}$ (and replacing the prefactor $|e_f H|/4 \pi^2$ with its continuum counterpart \cite{Chen_2016}). This step is called the continuum approximation. We get

\begin{eqnarray}
\frac{\partial \mathcal{F}^{\rm{med}}_{\rm{HLL}}}{\partial (\Delta^2)}\Big{|}_{\Delta = 0} = && \frac{N_c}{4 \pi^2} \sum_{f}  \bigintssss d p_{z} \bigintssss_{\sqrt{2 |e_f H|}}^{\infty} d p_{\perp} p_{\perp} \sum_{\zeta} \frac{1}{2 E^f_{\zeta}}  \nonumber \\
&& \times \bigg(1 + \zeta \frac{K}{\sqrt{p_z^2}}\bigg)\Theta\Big(\mu - E^f_{\zeta}\Big)\Big{|}_{\epsilon = 1} \ .
\label{eq:F_2}
\end{eqnarray}
We drop the index $n$ in $E^f_{\zeta}$ for the Landau levels when working in the continuum approximation. By carrying out the integration in $p_{\perp}$ and $p_z$ analytically, expanding for small magnetic fields (see \ref{sec:H_expansion}) and for small $K$, and summing the vacuum, medium and condensation contributions we reduce the qHGL expansion to

\begin{eqnarray}
\mathcal{F}_{\textrm{qHGL}}  = && \  \frac{1}{V}\int d\bold{x}\Big[ \mathcal{F}_{\hom}(\overline{M^2}, \mu, H) + \gamma_6\Big(3|\nabla M|^2|M|^2 \nonumber \\
&& + \frac{1}{2}(\nabla |M|^2)^2\Big) + \gamma_8\Big(14 |M|^4|\nabla M|^2 - \frac{1}{5}|\nabla M|^4  \nonumber \\
&& + \frac{18}{5}|M||\nabla^2 M||\nabla M|^2 +\frac{14}{5}|M|^2|\nabla^2M|^2\Big)\nonumber \\
&&  +  \sum_{m\geq 1}\tilde{\gamma}_{2m+2}|\nabla^mM|^2 \Big] \ .
\label{eq:qHGL}
\end{eqnarray}
The coefficients $\gamma_j$ (given in \ref{sec:Coefficients}) depend on the cutoff $\Lambda$, the chemical potential $\mu$ and on the magnetic field $H$. When the gradient terms in Eq. \eqref{eq:qHGL} multiply the magnetic-dependent part of the coefficients listed in \ref{sec:Coefficients}, one has to make the replacement \mbox{$\nabla \mapsto \bold{\hat{H}} \cdot \nabla$}, i.e. the gradients are calculated along the direction of the magnetic field. 

In the following we focus on the case $M(\bold{x}) = M(z)$, i.e. on field-aligned modulations. It has been shown for the CDW \cite{Frolov_2010} that a wave vector oriented perpendicularly to the magnetic axis is less favored energetically with respect to the solution with field-aligned wave vector. Additionally, although the approximation technique presented in this work can be extended to two-dimensional or three-dimensional modulations, no analytic expressions of the energy levels or spectral densities are known for two-dimensional or three-dimensional modulations, hindering the calculation of the free energy density of the LLL (which is the only Landau level not approximated by the qHGL functional) with the calculation technique adopted in this paper. One has to resort to the diagonalization of the infinite-dimensional hamiltonian matrix in momentum space \cite{Carignano_2012, Buballa_2015}, which is computationally demanding even for simple two-dimen-\\sional periodic crystalline structures.

By inspecting the coefficients $\gamma_j$, it is found that the gradient expansion converges as long as one has $\Delta, K, \sqrt{eH} < \mu$ \cite{Carignano_2018}. We relate the $\gamma_j$ and $\tilde{\gamma}_j$ as in \cite{Carignano_2018}, with $\gamma_4 = \tilde{\gamma_4}$, $\gamma_6 = 2 \tilde{\gamma_6}$ and $\gamma_8 = 5 \tilde{\gamma_8}$; for higher values of $j$ the relation has not been determined yet, and we set $\gamma_i = \tilde{\gamma_i}$ \cite{Carignano_2018}. 

In Eq. \eqref{eq:qHGL} the thermodynamic potential of homogeneous, magnetized quark matter $\mathcal{F}_{\hom}$ is given by 

\begin{eqnarray}
\mathcal{F}_{\hom}(\overline{M^2}, \mu, H) = \mathcal{F}_{\hom}^{\textrm{vac}} + \mathcal{F}_{\hom}^{\rm{med}} + \mathcal{F}_{\textrm{cond}} \ ,
\label{eq:F_hom_HLL}
\end{eqnarray}
where the vacuum and medium contributions read 

\begin{eqnarray}
\mathcal{F}_{\hom}^{\rm{vac}} = - \frac{N_c}{2 \pi^2} \sum_{f}\int_{\sqrt{2|e_fH|}}^{\infty}dp_{\perp} p_{\perp}\int dp_z E^f_{\textrm{PV}}
\end{eqnarray}

\begin{eqnarray}
\mathcal{F}_{\hom}^{\rm{med}} = &&   - \frac{N_c}{2 \pi^2}\sum_{f}\int_{\sqrt{2|e_fH|}}^{\infty}dp_{\perp} p_{\perp}\int dp_z \nonumber \\
&&  \times\Big[ (\mu - E^f)\theta(\mu - E^f)\Big] \ , 
\end{eqnarray}
where both the vacuum and medium contributions are calculated for $\epsilon = +$ and take into account the sum over the index $\zeta$, which yields a degeneracy factor.
Similarly to the IGL functional discussed in Section \ref{sec:IGL}, the $\mathcal{F}_{\hom}(\overline{M^2}, \mu, H)$ term in Eq.\eqref{eq:qHGL} reproduces the free energy of the homogeneous phase ($K = 0$) and of the long-wavelength phase ($K \ll \mu, \Delta$) of the quark condensate. In the presence of an external magnetic field, one also requires $\sqrt{eH} \ll \mu$ for the continuum approximation to be valid. The gradient terms in the first, second and third line of Eq. \eqref{eq:qHGL} are the gradients appearing in the standard Ginzburg-Landau approximation \cite{Nickel_2009, Abuki_2012, Carignano_2018}. The ones in the last line of Eq. \eqref{eq:qHGL} are obtained from the Taylor expansion in Eq. \eqref{eqn:Kappaexp}. They allow to determine with increased accuracy the condensate parameters using the qHGL functional with respect to the standard GL expansion close to the regions of the phase diagram where the second-order phase transition from the inhomogeneous, chirally broken phase to the chirally restored phase occurs (with $K \gg \Delta$ and $K \lesssim \mu$) \cite{Buballa_2015, Carignano_2018}. 

Similarly to the IGL approximation, the qHGL is valid for any periodic modulation.
The relation between the qHGL functional and the IGL approximation is

\begin{eqnarray}
\mathcal{F}_{\textrm{IGL}} = \lim_{H \to 0} \mathcal{F}_{\textrm{qHGL}} \ .
\label{eq:IGL_qHGL}
\end{eqnarray}
We note that the LLL contribution is proportional to $eH$ \cite{Frolov_2010, Cao_2016, Abuki_2018}, while the qHGL functional contains magnetic corrections to the quark free energy of higher orders.

\subsection{Nonzero temperatures}
\label{sec:qHGL_thermal}
High temperatures ($T \sim 10$ MeV) are plausible in astrophysical environments (such as neutron star mergers \cite{Radice_2018,Perego_2019, Endrizzi_2020}), can be probed in laboratories \cite{Jacobs_2005,Adams_2005}, and can affect the phase diagram of quark matter \cite{Nakano_2005, Nickel_2009, Buballa_2015}. 

For $T >0$, the quark free energy density (Eq. \eqref{eq:F_MAG}) acquires an additional contribution $\mathcal{F}_{\textrm{T}}$, viz.

\begin{eqnarray}
\mathcal{F} = \mathcal{F}_{\textrm{LLL}} + \mathcal{F}_{\textrm{qHGL}} + \mathcal{F}_{\textrm{T}} \ .
\label{eq:F_MAG_T}
\end{eqnarray}
The thermal contribution $\mathcal{F}_{\textrm{T}}$ reads \cite{Frolov_2010}

\begin{eqnarray}
\mathcal{F}_{\textrm{T}}  = && - N_c \beta^{-1}\sum_{f}\frac{|e_f H| }{(2\pi)^2}  \int dp_z \sum_{n, \zeta, \epsilon} \log\Big(1 + e^{-\beta |E-\mu|}\Big)  \nonumber \\
&& 
\label{eq:thermal_contribution}
\end{eqnarray}
with $\beta = T^{-1}$, where we denote with $E$ the eigenvalues given in Eq. \eqref{eq:energy_levels} for simplicity. As for the $T = 0$ case presented in Section \ref{sec:qHGL}, the thermal contribution can be divided into the LLL ($n = 0$) and higher Landau levels ($n > 0$) contributions (which we denote with $\mathcal{F}_{\textrm{T,LLL}}$ and $\mathcal{F}_{\textrm{T,HLL}}$ respectively). For sufficiently weak magnetic fields ($\sqrt{eH} \ll \mu$), we approximate the sum over the index $n$ labeling the higher Landau levels in $\mathcal{F}_{\textrm{T,HLL}}$ with an integration over the quark momenta perpendicular to the magnetic field, i.e.

\begin{eqnarray}
\mathcal{F}_{\textrm{T,HLL}}  = && - \frac{N_c \beta^{-1} }{4\pi^2} \sum_{f}\int d p_z \int_{\sqrt{2 |e_f H|}}^{\infty} d p_{\perp} p_{\perp} \nonumber \\ 
&& \times \sum_{\zeta, \epsilon} \log\Big(1 + e^{-\beta |E-\mu|}\Big) . 
\label{eq:thermal_contribution_HLL_0}
\end{eqnarray}
It is possible to approximate Eq. \eqref{eq:thermal_contribution_HLL_0} with a procedure similar to the one presented in \cite{Cao_2016}. One can calculate the $\mathcal{F}_{\textrm{T,HLL}}$ contribution introducing the density of states $\rho$ \cite{Buballa_2015}, which is a simpler form to expand $\mathcal{F}_{\textrm{T,HLL}}$ in powers of $\Delta$ and $K$. The result is identical to Eq. \eqref{eq:thermal_contribution_HLL_0}.

\section{Phase diagram at zero temperature}
\label{sec:QCD_phase_diagram}
In this section we investigate the phase diagram for inhomogeneous, magnetized quark matter at zero temperature. The effect of magnetic fields on inhomogeneous chiral condensates has been investigated in numerous works \cite{Frolov_2010, Nishiyama_2015, Carignano_2015, Buballa_2016, Abuki_2018}. Differently from most of the literature, we study the condensate parameters in the presence of an external field employing the Pauli-Villars regularization scheme, which confers certain practical calculational advantages discussed above.

We assume that the inhomogeneous quark condensate is modulated in two different ways: as a CDW (Eq. \eqref{eq:CDW}), or as a solitonic modulation (\enquote{real kink crystal}, RKC) given by

\begin{eqnarray}
M_1(z) = \Delta_{1}\sqrt{\nu} \  \textrm{sn}(\Delta_1 z | \nu) \ ,
\label{eq:soliton}
\end{eqnarray}
where $\textrm{sn}(\Delta_1 z | \nu)$ is the Jacobi elliptic sine and $\nu$ is the elliptic modulus \cite{Nickel_2009, Buballa_2015}. The shape of the modulation depends on $\nu$. For $\nu \rightarrow 1$, the elliptic sine coincides with the hyperbolic tangent; as $\nu$ approaches zero, the mass function approaches a sinusoidal modulation. It has been suggested \cite{Bashar_2009, Nishiyama_2015} that in magnetized quark matter the \enquote{hybrid chiral condensate} (given by Eq. \eqref{eq:soliton} times a complex phase) is thermodynamically favored. In this work we do not consider for simplicity the hybrid chiral condensate and focus for definiteness on Eqs. \eqref{eq:CDW} and \eqref{eq:soliton}.

The plan of this section is as follows. In Section \ref{sec:Benchmark_Mag} we study the CDW modulation, comparing the unmagnetized and magnetized cases and relating our results with previous ones obtained in the literature \cite{Frolov_2010, Carignano_2015, Buballa_2016}. Following this benchmarking exercise, in  Section \ref{sec:Solitons_Mag} we generalize the qHGL method to other, non-plane-wave modulations of the quark condensate. In the following, we fix $\Lambda = 757.048$ MeV and $G_S \Lambda^2 = 6.002$ \cite{Carignano_2018}, which lead to an effective quark mass in vacuo of $300$ MeV and a pion decay constant of $f_{\pi} = 88$ MeV.

\subsection{Benchmarking the qHGL expansion}
\label{sec:Benchmark_Mag}

\begin{figure*}
  \includegraphics[width=17.5 cm, height = 6.3 cm]{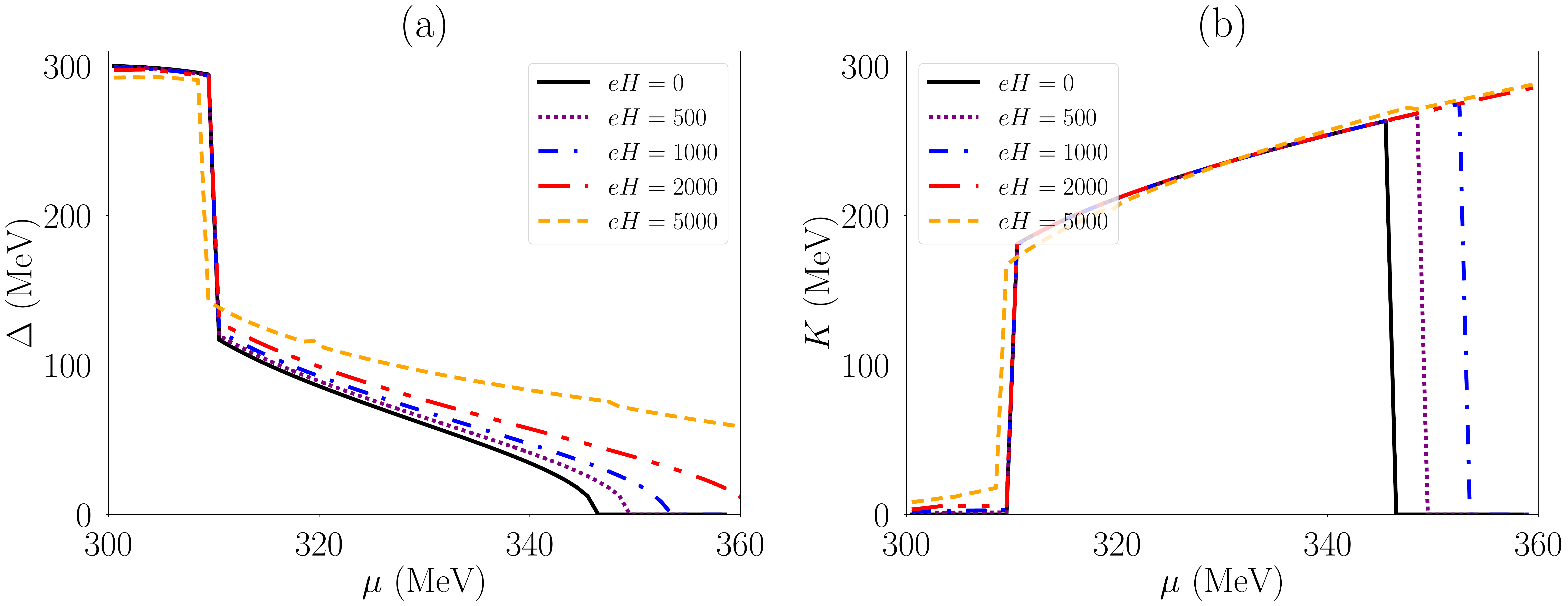}
\caption{CDW parameters for $0 \leq eH \leq 5000$ MeV$^2$. In this work we use the Pauli-Villars regularization, which confers calculational advantages and supplements analogous calculations with proper-time regularization in the literature. \textit{(a)} Amplitude $\Delta$ as a function of the chemical potential. \textit{(b)} Parameter $K$ versus $\mu$. The legends label the magnetic field strength $eH$ in units of MeV$^2$.}
\label{fig:qHGL}       
\end{figure*}

The thermodynamically favored state of the unmagnetized quark condensate is found by minimizing the free energy density with respect to the amplitude and wave vector of the trial modulation for fixed chemical potential. We study first the CDW modulation in the presence of an external magnetic field as a benchmark model for the qHGL functional. We minimize the quark free energy in Eq. \eqref{eq:F_MAG}, with $\mathcal{F}_{\textrm{LLL}}$ calculated similarly to \cite{Frolov_2010}, while the free energy density of the higher Landau levels is approximated by the qHGL functional truncated at $\gamma_{10}$. The results are obtained using Pauli-Villars regularization, which confers calculational advantages discussed above and supplements analogous calculations with proper-time regularization in the literature \citep{Frolov_2010}. 

Figure \ref{fig:qHGL}(a) shows the amplitude of the CDW as a function of the quark chemical potential $\mu$ for different values of $eH$ (in units of MeV$^2$). For $eH = 0$ (black, solid line) and $\mu \lesssim 310.5$ MeV, the quark condensate is homogeneous, with $\Delta \approx 300.0$ MeV and $K = 0$. At $\mu \approx 310.5$ MeV there is a first-order phase transition to the inhomogeneous phase, in which the amplitude $\Delta$ falls to  $\Delta \approx 117.0$ MeV, while $K$ jumps from $K = 0$ to $K \approx 180.4 $ MeV. Chiral restoration ($\Delta, K = 0$) occurs at $\mu_c \approx 346.5$ MeV. Upon including higher order gradients (proportional to $\gamma_{12}$ for example), the value of $\mu_c$ is slightly lower, with $\mu_c \approx 345.5$ MeV. For $eH = 500$ MeV$^2$ (purple, dotted line) and $\mu < 310.5$ MeV, $\Delta$ is degenerate with the unmagnetized case, while for $\mu \geq 310.5$ MeV the blue curve is slightly higher than the black curve. The chiral condensate melts ($\Delta, K = 0)$ at $\mu_c \approx 346.5$ MeV for $eH = 0$ and at $\mu_c \approx 349.5$ MeV for $eH = 500$ MeV$^2$, i.e. the magnetic field shifts the transition from the inhomogeneous phase to the restored one of $\approx 3.0$ MeV with respect to the unmagnetized case, enlarging the region where the condensate is inhomogeneous with $K > \Delta$ ($\mu \in [310.5, 346.5]$ MeV for $eH = 0$) by $ \approx 8$ per cent.

By further increasing $eH$, the amplitude of the chiral density wave becomes larger than the unmagnetized case, and the chiral restoration point gradually shifts to higher chemical potentials, in line with previous results \cite{Frolov_2010, Carignano_2015, Buballa_2016}. For $eH = 1000$ MeV$^2$ (blue, dashed-dotted curve) the condensate vanishes at $\mu_c \approx 353.5$ MeV. For stronger magnetic fields ($eH \gtrsim 2000$ MeV$^2$) the continuum approximation for the higher Landau level becomes less accurate, as the condition $\sqrt{eH} \ll \mu$ is not strictly satisfied. For $eH = 2000$ MeV$^2$ (red, broken curve) the amplitude $\Delta$ vanishes for $\mu \approx 361.5$ MeV, while for $eH = 5000$ MeV$^2$ (orange, dashed line) the first-order phase transition for the orange curve occurs at $\mu \approx 309.5$ MeV and $\Delta >0$ for $\mu > 370.0$ MeV, in line with previous results \cite{Buballa_2016}. For magnetic fields typical of highly-magnetized neutron stars (with $H \approx 1.4 \times 10^{18}$ G in the core), it has been shown \cite{Frolov_2010, Carignano_2015, Buballa_2016} that the quark condensate does not melt up to high values of the chemical potential ($\mu \approx 500$ MeV).

In Figure \ref{fig:qHGL}(b) the parameter $K$ of the CDW modulation is displayed as a function of $\mu$ for various magnetic field strengths. The results are similar to the unmagnetized case, but as shown in Figure \ref{fig:qHGL}(a) the transition to the chirally restored phase ($\Delta, K = 0$) shifts to higher chemical potentials with respect to $eH = 0$. For weak magnetic fields ($eH \leq 500$ MeV$^2$) we find $0 < K \lesssim 1.5 $ MeV even for small chemical potentials ($\mu \lesssim 310.5$ MeV), while for stronger fields ($500 \ \textrm{MeV}^2 \lesssim eH \leq 5000$ MeV$^2$) we obtain $1.5 \ \textrm{MeV} \lesssim K \lesssim 20$ MeV. The first order phase transition (with $K > \Delta$) occurs at $\mu \approx 310.5$ MeV for $eH \leq 2000$ MeV$^2$ and at $\mu \approx 309.5$ MeV for $eH = 5000$ MeV$^2$. In the presence of a magnetic field the inhomogeneous phase extends to $\mu > 0$, contrarily to the unmagnetized case, as found in \cite{Frolov_2010, Carignano_2015, Buballa_2016}. This effect is due to the spectral asymmetry of the LLL, which leads to the presence of an anomalous term proportional to $-|eH| \mu K$ \cite{Frolov_2010} in the free energy density $\mathcal{F}_{\textrm{LLL}}$, which favors the formation of an inhomogeneous chiral condensate for arbitrary  positive values of $\mu$.

The extension of the inhomogeneous phase to a larger region of the phase diagram is driven mainly by the LLL, which gives a contribution to the free energy density of order $eH$. The higher Landau levels contribute with magnetic corrections and favor the enlargement of the region where the condensate is inhomogeneous (although less than the LLL). The magnetic-field-dependent coefficients of the qHGL functional slightly increase the free energy gain for $H, \Delta, K \neq 0$, and in particular favor the inhomogeneous phase close to $\mu_c$ where the gradients of the condensate are large.

How accurate is the qHGL expansion? We examine this important question in \ref{sec:Comparison}. We compare quantitatively the CDW parameters calculated with the qHGL expansion and with the non-approximated free energy. We find that for $\mu \lesssim 340$ MeV the condensate parameters are determined to an accuracy of $\lesssim 13$ per cent. We also find a qualitative agreement with the literature, where the proper-time regularization scheme is often employed. We find for example that chiral restoration is suppressed for strong magnetic fields up to $\mu \approx 400$ MeV. The accuracy of the qHGL expansion will be improved in future work.

In summary, the qHGL expansion is a practical method that approximates the free energy of the CDW in a magnetic field. It can be applied throughout the phase diagram of magnetized quark matter at $T = 0$, unlike the standard GL expansion (which cannot be applied reliably at $T = 0$ and which is limited to the regions of the phase diagram with $T > 0$ where the amplitude of the condensate and its gradients are small \cite{Buballa_2015, Abuki_2018}). The qHGL expansion simplifies previous numerical methods that employ different regularization schemes for the evaluation of the thermodynamic potential \cite{Klevansky_1992, Frolov_2010, Carignano_2015} and allows to compute quickly the combinations ($\Delta$, $K$) including the values where first-order phase transition and chiral restoration occur. The results reported in Figure \ref{fig:qHGL} agree with previous numerical studies \cite{Frolov_2010, Carignano_2015, Buballa_2016} (although we use a different regularization scheme), showing that an external magnetic field favors the inhomogeneous chirally broken phase ($\Delta, K \neq 0$) which extends to small chemical potentials ($\mu > 0$) and above $\mu_c \approx 346.5$ MeV (the chiral restoration point in unmagnetized quark matter). 

We discuss briefly possible extensions of the method developed in this work, namely the conditions of $\beta$-equilibrium (with electrical neutrality) and the case of nonvanishing quark bare masses in \ref{sec:Applications}.

\subsection{Solitons in magnetized quark matter}
\label{sec:Solitons_Mag}

\begin{figure*}
  \includegraphics[width=17.5 cm, height = 6.3 cm]{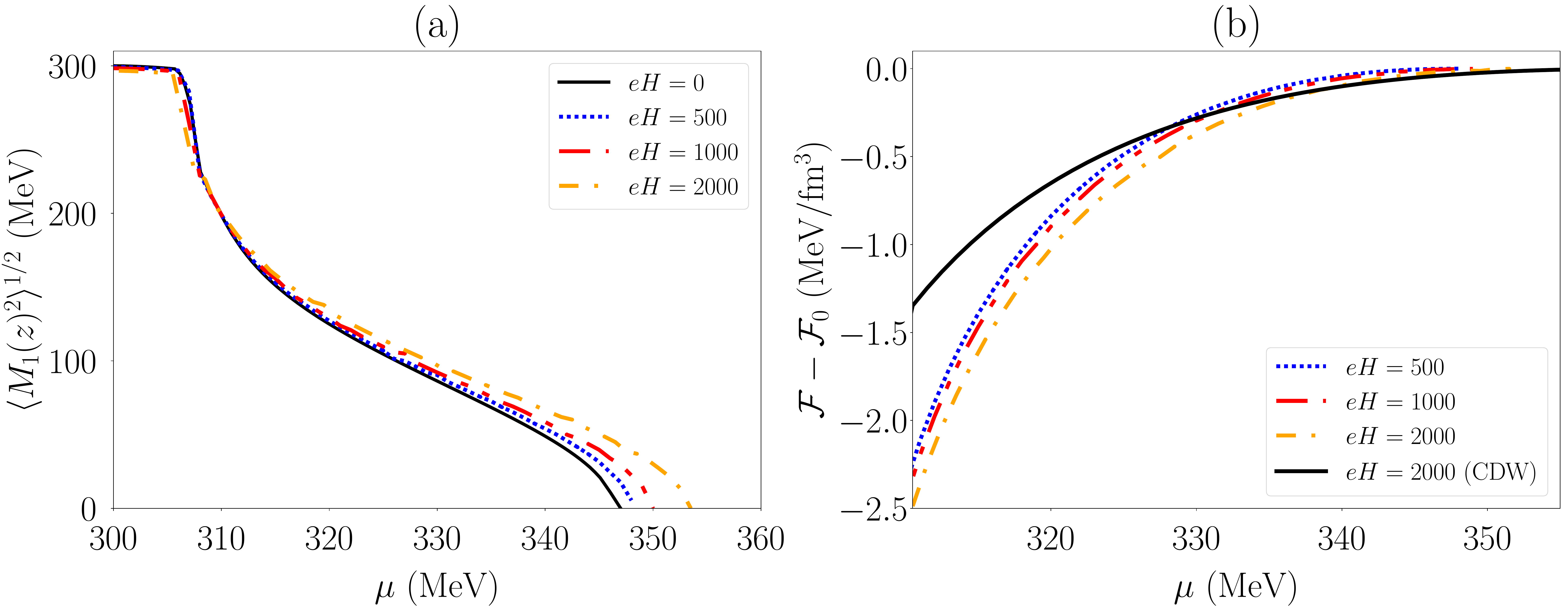}
\caption{RKC modulation and modification of the ground state of inhomogeneous quark matter due to magnetic fields for $eH \in [0, 2000]$ MeV$^2$. \textit{(a)} Cell-averaged value of the effective mass. \textit{(b)} Free energy gain for the RKC condensate and for the CDW. The legends report the magnetic field in units of MeV$^2$.}
\label{fig:qHGL_SOL}       
\end{figure*}

The RKC modulation is thermodynamically favored over the CDW for $eH = 0$, as it produces the larger free energy gain \cite{Nickel_2009, Buballa_2015, Carignano_2018}. On the other hand, it has been shown \cite{Bashar_2009, Frolov_2010, Nishiyama_2015, Abuki_2018} that a sufficiently strong magnetic field favors the formation of condensates with a complex phase as the CDW over the RKC modulation.
In this section we employ the qHGL approximation to study the RKC condensate in the presence of a magnetic field. This generalizes the treatment of the CDW in Section \ref{sec:Benchmark_Mag}. As for the CDW, the RKC condensate in the presence of a magnetic field is often studied with the proper-time regularization scheme, or employing Ginzburg-Landau techniques \cite{Cao_2016}.
We show that the qHGL expansion extends previous studies employing a standard GL expansion beyond the region of the phase diagram where the amplitude of the condensate and its gradients are small \cite{Cao_2016}. We also show that our method agrees qualitatively with previous numerical studies obtained with different regularization schemes \cite{Nishiyama_2015}. The free energy density is given by the sum of the LLL contribution and the qHGL functional. For the RKC modulation, we follow \cite{Cao_2016} and use the density of states reported in \cite{Nickel_2009, Cao_2016} to calculate the $\mathcal{F}_{\textrm{LLL}}$ term in Eq. \eqref{eq:F_MAG}. 

Figure \ref{fig:qHGL_SOL}(a) shows the cell-averaged value of the effective mass for the RKC modulation $\langle M_1(z)^2 \rangle^{1/2}$ for values of the magnetic field in the range $eH \in [0, 2000]$ MeV$^2$. First we briefly review the unmagnetized case \cite{Nickel_2009,Buballa_2015, Carignano_2018} as a benchmarking exercise, to be compared with the magnetized case. For $eH = 0$ and for $300.0 \ \textrm{MeV} \leq \mu \lesssim 307.0$ MeV the homogeneous phase is characterized by $\langle M_1(z)^2\rangle^{1/2} \approx 300.0$ MeV. The transition from the homogeneous to the inhomogeneous phase is of second order, with $\langle M_1(z) \rangle^{1/2}$ decreasing smoothly at $\mu \gtrsim 307.0$ MeV and approaching zero at $\mu_c \approx 346.5$ MeV. For $eH = 500$ MeV$^2$ and $eH = 1000$ MeV$^2$ the cell-averaged value of the condensate decreases smoothly from $ \langle M_1(z)^2 \rangle^{1/2} \approx 300.0$ MeV for $\mu \approx 307.0$ MeV, while for $eH = 2000$ MeV$^2$ the transition occurs at $\mu \approx 306.5$ MeV. The phase transition to the inhomogeneous phase shifts to lower chemical potentials for $eH = 2000$ MeV$^2$ for the RKC modulation, while for the CDW the shift of the (first-order) phase transition takes place for $eH = 5000$ MeV$^2$. The condensate melts at $\mu_c \approx 349.0$ MeV for the blue, dotted line ($eH = 500$ MeV$^2$), at $\mu_c \approx 350.0$ MeV for the red, broken line ($eH = 1000$ MeV$^2$) and at $\mu_c \approx 353.0$ MeV for the orange, dashed-dotted line ($eH = 2000$ MeV$^2$). The different curves in Figure \ref{fig:qHGL_SOL}(a) are similar. Contrarily to the CDW modulation (Figure \ref{fig:qHGL}(a)), the magnetic field has a small effect on the phase diagram of the RKC (in agreement with the previous numerical studies performed with different regularization schemes \cite{Nishiyama_2015}), with a shift of $\mu_c$ to higher values of just $6.5$ MeV for $eH = 2000$ MeV$^2$ with respect to the unmagnetized case (black, solid curve). We emphasize that the results reported in Figure \ref{fig:qHGL_SOL}(a) are calculated using the coefficients up to $\gamma_{10}$ in the qHGL expansion. By including higher order corrections, we find for example that the chiral restoration is shifted by only $\approx 1.5$ MeV for $eH = 1000$ MeV$^2$ with respect to the unmagnetized case (not shown here), in line with previous results \cite{Nishiyama_2015}. 

Figure \ref{fig:qHGL_SOL}(b) displays the free energy gain for the RKC modulation for $500 \ \textrm{MeV}^2 \leq eH \leq 2000 \ \textrm{MeV}^2$ (blue, red and orange curves) and for the CDW modulation (black curve) for $eH = 2000$ MeV$^2$. As the magnetic field strengthens, the free energy gain becomes larger. For $\mu \lesssim 335$ MeV the orange curve lies below the black one, showing that in moderately magnetized quark matter the solitonic shape of the chiral condensate is favored. For $\mu \gtrsim 335$ MeV, the plane-wave modulation produces a larger free energy gain than the RKC shape, and the ground state of magnetized quark matter is the CDW \cite{Frolov_2010, Nishiyama_2015, Abuki_2018}, contrarily to the unmagnetized case. The stability of the modulated condensate for strong magnetic fields is related to the spectral asymmetry of the LLL and is therefore different for different modulations, e.g. CDW versus solitonic. For the CDW, the spectrum of the LLL is asymmetric. The asymmetry originates from the chiral anomaly due to the coupling of the magnetic field with the axial-vector current (and hence with the phase of the CDW, cf. \cite{Tatsumi_2015, Abuki_2018}). As a consequence of the spectral asymmetry, the LLL contribution of the CDW to the free energy has an additional anomalous term $-|eH| \mu K$ \cite{Frolov_2010}. In contrast, there is no chiral anomaly in the solitonic modulation for example (which does not depend on a complex phase), and the corresponding LLL contribution does not include the anomalous term $-|eH| \mu K$. These features have consequences for the CDW condensate parameters and the phase diagram of magnetized quark matter, stabilizing the CDW with respect to the solitonic modulation provided that the magnetic field is sufficiently high.

In summary, we verify for the RKC modulation that the qHGL expansion agrees with previous works employing the proper-time regularization scheme \citep{Nishiyama_2015}. It is a valid technique to calculate $\langle M_1(z)^2 \rangle^{1/2}$ and thereby infer the order of the phase transition and free energy gain of the chiral condensate at $T = 0$. It can be applied throughout the phase diagram of magnetized quark matter extending previous studies employing GL expansion techniques \cite{Cao_2016}. It predicts that the magnetic field favors the plane-wave modulation over the RKC modulation at high densities ($\mu \gtrsim 335$ MeV) modifying the ground state of magnetized quarks \cite{Nishiyama_2015}.

\begin{figure*}
  \includegraphics[width=16.5 cm, height = 12.3 cm]{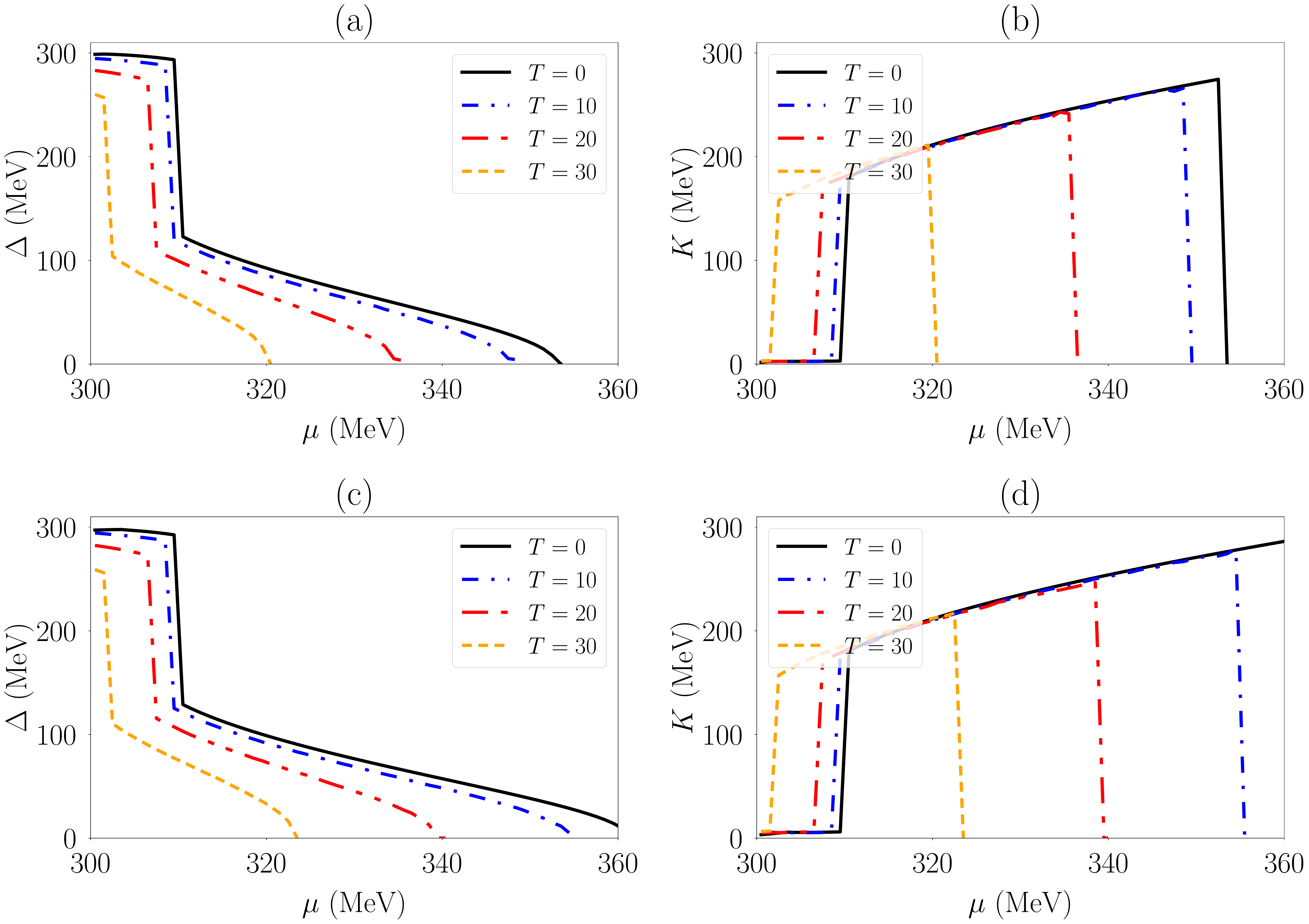}
\caption{CDW parameters for two typical magnetic field strengths ($eH = 1000$ MeV$^2$ and $eH = 2000$ MeV$^2$) at different temperatures. \textit{(a)} Amplitude $\Delta$ versus $\mu$ for $eH = 1000$ MeV$^2$. \textit{(b)} Parameter $K$ as a function of $\mu$ for $eH = 1000$ MeV$^2$. \textit{(c) and (d)} As panels (a) and (b) respectively, but for $eH = 2000$ MeV$^2$. The legends report the temperature in units of MeV.}
\label{fig:qHGL_CDW_T}       
\end{figure*}

\section{Nonzero temperatures}
\label{sec:QCD_phase_diagram_T}

In this section we consider the case of magnetized quarks with $T > 0$. Temperatures of the order of $T \approx 10$ MeV may have important applications in binary mergers of neutron stars for example \cite{Radice_2018,Perego_2019, Endrizzi_2020}) and are in the reach of upcoming laboratory experiments \cite{Jacobs_2005,Adams_2005}. Previous studies of magnetized, hot matter focus on the phase diagram of quark matter close to the Lifshitz point ($T \approx T_c$) \cite{Tatsumi_2015, Abuki_2018}. We focus on the regime $0 \lesssim T \lesssim 30$ MeV and $T \sim\sqrt{eH} \lesssim 45$ MeV, which is lightly explored in the literature. In this regime, thermal and magnetic effects are comparable ($T \sim\sqrt{eH}$) and compete to set the condensate amplitude and the size of the density region where the inhomogeneous phase is favored. For simplicity, we limit ourselves to the CDW modulation. The study of different modulations for $T, H > 0$ is left for future work.

Figure \ref{fig:qHGL_CDW_T} reports the amplitude $\Delta$ and parameter $K$ of the CDW for temperatures in the range $0 \leq T \leq 30$ MeV for $eH = 1000$ MeV$^2$ and $eH = 2000$ MeV$^2$. In Figure \ref{fig:qHGL_CDW_T}(a) we study how thermal and magnetic effects compete in setting the condensate amplitude. As $T$ increases, the CDW amplitude for $eH = 1000$ MeV$^2$ decreases throughout the region $300.0 \ \textrm{MeV} \lesssim \mu \lesssim 354.0$ MeV. The region in which the chiral condensate is inhomogeneous shrinks with respect to $T = 0$. For $T = 10$ MeV (blue, dotted-dashed line), the first-order phase transition from the homogeneous to the inhomogeneous phase occurs at $\mu \approx 309.5$ MeV, with the CDW amplitude attaining $\Delta = 288.0$ MeV at $\mu = 308.5$ MeV and $\Delta = 119.5$ MeV at $\mu = 309.5$ MeV. The condensate melts at $\mu_c \approx 349.0$ MeV, contrarily to the black, solid curve ($T = 0$, with $\mu_c \approx 353.5$ MeV). Similarly for $T = 20$ MeV (red, broken curve), $\Delta$ jumps discontinuously at $\mu \approx 307.5$ MeV, and chiral restoration is reached at $\mu_c \approx 336.0$ MeV. For higher temperatures ($T = 30$ MeV, orange dashed line), the first-order phase transition of $\Delta$ occurs at $\mu \approx 302.5$ MeV, with $\Delta = 104.0$ MeV. The condensate melts at $\mu_c \approx 320.5$ MeV.

Figure \ref{fig:qHGL_CDW_T}(b) displays the corresponding parameter $K$ for $eH = 1000$ MeV$^2$. It shows that the density region where the condensate is inhomogeneous (with $K > \Delta$) is shrunk by thermal effects. For $T > 0$ MeV, $K$ jumps at the first-order transition up to $K \approx 176.1$ MeV for $T = 10$ MeV (blue, dotted-dashed curve), to $K \approx 171.3$ MeV for $T = 20$ MeV (red, broken curve) and to $K \approx 157.7$ MeV for $T = 30$ MeV (orange, dashed curve). At higher chemical potentials, $K$ plummets to $K = 0$ where the amplitude of the condensate vanishes (cf. Figure \ref{fig:qHGL_CDW_T}(a)). By comparing Figure \ref{fig:qHGL_CDW_T}(b) with the $eH = 0$ case (Figure 6 in \cite{Buballa_2015}), we see that the region of the phase diagram where the condensate is inhomogeneous is larger for $eH \neq 0$, as the magnetic field favors chiral symmetry breaking, in competition with thermal effects.

We study the condensate parameters for $T > 0$ and $eH = 2000$ MeV$^2$ in Figures \ref{fig:qHGL_CDW_T}(c) and \ref{fig:qHGL_CDW_T}(d). The main trend in Figure \ref{fig:qHGL_CDW_T}(c) is that the stronger magnetic field increases $\Delta$ with respect to Figure \ref{fig:qHGL_CDW_T}(a), while thermal effects tend to reduce $\Delta$. For $T = 10 $ MeV (blue, dotted-dashed line), the CDW amplitude $\Delta$ jumps discontinuously at $\mu \approx 309.5$ MeV, attaining $\Delta = 125.3$ MeV. Chiral restoration takes place at $\mu_c \approx 355.0$ MeV while for $T = 0$ $\mu_c \approx 361.5$ MeV. For $T = 20$ MeV (red, broken curve), we find $\Delta \approx 115.4$ MeV at $\mu \approx 307.5$ MeV, and $\Delta = 0$ at $\mu_c \approx 340.0$ MeV. Thermal effects are stronger for $T = 30$ MeV (orange dashed line), the first-order phase transition occurs at $\mu \approx 302.5$ MeV, and the condensate melts at $\mu_c \approx 323.5$ MeV.

Figure \ref{fig:qHGL_CDW_T}(d) reports the parameter $K$ for $eH = 2000$ MeV$^2$. The density region where the condensate inhomogeneous with $K > \Delta$ is larger than Figure \ref{fig:qHGL_CDW_T}(b) due to the stronger magnetic field. For $T = 10$ MeV (blue, dotted-dashed curve) $K$ jumps up to $K \approx 175.8$ MeV at $\mu = 309.5$ MeV . For $T = 20$ MeV (red, broken curve) the $K$ parameter increases up to $K \approx 171.0$ MeV at $\mu = 307.5$ MeV, and to $K \approx 157.1$ MeV at $\mu = 302.5$ MeV for $T = 30$ MeV (orange, dashed curve). When chiral restoration is reached (cf. Figure \ref{fig:qHGL_CDW_T}(c)), we find $K = 0$.

Unlike the standard GL approach \cite{Cao_2016, Tatsumi_2014, Abuki_2018}, which is valid only close to the Lifshitz point (i.e. where both the amplitude of the condensate and its gradients are small), the gradient expansion technique developed in this work is valid throughout the $\mu$-$T$ plane. The qHGL functional simplifies the evaluation of the non-thermal contribution to the thermodynamic potential with respect to previous numerical approaches \citep{Nishiyama_2015, Carignano_2015}, allowing to determine the chiral condensate parameters, phase transition and critical restoration points away from the Lifshitz point in the $\mu $-$T$ plane. We find that temperatures that may be plausible for newly born neutron stars (for example in binary mergers \cite{Radice_2018, Perego_2019, Endrizzi_2020}) and that are in the reach of upcoming laboratory experiments \cite{Jacobs_2005, Adams_2005} have a significant effect on the phase diagram of magnetized quark matter.

\section{Conclusion}
Magnetic fields in the cores of standard neutron stars and magnetars may reach $H \approx 10^{16}$ G and $H \approx 10^{18}$ G respectively \cite{Lyne_2006, Ferrer_2010, Carignano_2015, Buballa_2016}, which can affect the ground state of deconfined quarks. The phase diagram of magnetized quark matter in the presence of an external magnetic field is studied using an analytic Ginzburg-Landau expansion (qHGL) of the free energy density of magnetized, inhomogeneous quark matter. The qHGL functional extends previous results \cite{Carignano_2018} for inhomogeneous, unmagnetized quarks and is valid for any periodic modulation of the quark condensate. Compared to previous studies employing standard GL techniques (limited to the regions of the quark phase diagram where both the amplitude of the chiral condensate and its gradients are small), the qHGL expansion is valid throughout the phase diagram of magnetized quark matter, and provides an analytic approximation of the quark free energy that simplifies previous numerical studies \cite{Buballa_2015, Carignano_2015, Nishiyama_2015, Abuki_2018}. It is easy to evaluate the qHGL expansion quickly.

Inhomogeneous phases of magnetized quark matter are studied often with the proper-time regularization scheme \cite{Frolov_2010, Nishiyama_2015, Buballa_2015}. In this work we study the chiral condensate parameters using the Pauli-Villars regularization scheme, which confers calculational advantages as discussed above. At zero temperature, the free energy density is given by the sum of the contribution of the lowest Landau level \cite{Landau_1981, Frolov_2010} and the qHGL functional. For a plane-wave ansatz, as a benchmarking exercise, we find that the presence of an external magnetic field favors the inhmomogeneous, chirally broken phase of the quark condensate in line with previous works \cite{Frolov_2010, Carignano_2015, Buballa_2016, Abuki_2018}. The unmagnetized quark condensate vanishes at the chiral restoration point, located at $\mu_c \approx 346.5$ MeV. The magnetic field shifts the chiral restoration point to higher chemical potentials. It also favors the formation of an inhomogeneous quark condensate for $\mu > 0$ due to the spectral asymmetry of the LLL \cite{Frolov_2010}. The amplitude and wave vector of the plane-wave modulation vanish at $\mu_c \approx 349.5$ MeV for $eH = 500$ MeV$^2$, at $\mu_c \approx 353.5$ MeV for $eH = 1000$ MeV$^2$ and at $\mu_c \gtrsim 361.5$ MeV for $eH \gtrsim 2000$ MeV$^2$. We test the accuracy of the qHGL approximation by comparing with the non-approximated result $\mathcal{F}_{\textrm{num}}$ (see \ref{sec:Comparison}). We find that the CDW amplitude is determined to within $\lesssim 13$ per cent for $\mu \lesssim 340$ MeV, while $K$ to within $\lesssim 2$ per cent.

The qHGL approximation does not apply only to plane-wave modulations. It can be applied readily to solitonic modulations as well for example. When plane-wave and solitonic modulations are compared, we find that the latter is less influenced than the former by the magnetic field \cite{Nishiyama_2015} for $0 \lesssim eH \lesssim 2000$ MeV$^2$. For the solitonic condensate, the shift to lower chemical potential of the second-order phase transition to the inhomogeneous phase happens already at $eH = 2000$ MeV$^2$. On the other hand, the chiral restoration point is located at $\mu_c \approx 353.0$ MeV for $eH = 2000$ MeV$^2$. The qHGL expansion predicts that, for sufficiently strong magnetic fields ($eH \approx 2000$ MeV$^2$) and for $\mu \gtrsim 335$ MeV, the thermodynamically favored ground state of inhomogeneous quarks is a plane wave, in line with previous studies \cite{Nishiyama_2015, Abuki_2018}.

For $T > 0$ and $H > 0$, the qHGL functional approximates the non-thermal behavior of magnetized quark matter. We focus on the experimentally relevant case $0 \lesssim T \lesssim 30$ MeV and $T \sim \sqrt{eH} \lesssim 45$ MeV, i.e. where the thermal energy scale is comparable to the separation of the Landau levels. In this regime, thermal and magnetic effects compete in setting the condensate parameters and the density range of the inhomogeneous phase. Astrophysical applications with $T \approx 30$ MeV and $eH \lesssim 2000$ MeV$^2$ remain lightly explored in the literature. When the chiral condensate is modulated as a plane wave, we find that the first-order phase transition from the low-frequency phase (with $K < 20$ MeV) to the high-frequency phase (with $K \gtrsim 150$ MeV) shifts to lower chemical potentials, and that the region where the inhomogeneous phase is favored is shrunk by nonzero temperatures \cite{Nakano_2005, Buballa_2015}. For example, the interval in $\mu$ where the condensate is inhomogeneous for $T = 30$ MeV and $eH = 2000$ MeV$^2$ is only 40 per cent of the corresponding $\mu$ interval obtained for $T = 0$ and $eH = 2000$ MeV$^2$. By contrast, the interval in $\mu$ increases, when $T$ is fixed and $H$ increases. Hence, the qHGL expansion is a tool to explore the phase diagram of hot and magnetized quark matter (cf. standard GL approximations, which are valid near the Lifshitz point), i.e. conditions that may be found in neutron stars mergers \cite{Radice_2018,Perego_2019, Endrizzi_2020}) and that can be probed by laboratory experiments \cite{Jacobs_2005,Adams_2005}.

In summary, this work shows that the qHGL functional offers a practical route to calculating the free energy density and condensate parameters of magnetized, inhomogeneous quark matter. It can be applied to arbitrary periodic modulations of the chiral condensate, not just the CDW, and to calculate magnetic corrections in the intermediate-field regime $H \lesssim 10^{18}$ G, extending standard Ginzburg-Landau techniques away from the Lifshitz points. The choice of the Pauli-Villars regularization enables the computational advantages shown in this paper, simplifying the study of inhomogeneous phases both in cold ($T = 0$) and hot ($T > 0$) quark matter in the presence of a magnetic field. As a demonstration of how it can be applied, the qHGL approximation is used at zero temperature to show that solitonic condensates are favored over the CDW for $H \lesssim 10^{16}$ G, while on the contrary for $H \approx 10^{17}$ G the CDW supersedes the soliton shape for $\mu \gtrsim 335$ MeV. For $T >0 $, the qHGL approximation allows to study easily temperature and magnetic field regimes typical of astrophysical environments, and correctly predicts that the inhomogeneous phase shifts to lower chemical potentials, reducing considerably the size of the region where the condensate is inhomogeneous.

We conclude by summarising the practical applicability of the calculations in this paper in the limit $T \rightarrow 0$. If one is interested in studying the properties of cold ($T \ll 10^{11}$ K), magnetized neutron star cores with $H \lesssim 10^{16}$ G, the unmagnetized limit of the qHGL is sufficient to describe the ground state of quark matter. If on the other hand one is interested in stronger magnetic fields ($ 10^{16} \ \textrm{G} \lesssim H \lesssim  10^{18}$ G), the qHGL is a practical technique to determine the parameters of the chiral condensate. As an example of why this is useful, the neutrino emissivity produced by the quark beta decay depends on the amplitude and wave vector of the chiral condensate. It has been calculated in the case of magnetized, homogeneous \cite{Xue_Wen_2007} and unmagnetized, inhomogeneous quarks \cite{Tatsumi_2014}. It is stronger than modified Urca processes in nuclear matter \cite{Yakovlev_2001} and can provide a signature of the presence of quark matter in neutron stars. In particular, the emissivity of unmagnetized, inhomogeneous quarks (calculated for a CDW modulation) has been evaluated near the onset of the inhomogeneous phase and of the chiral restoration transition, which are determined with good accuracy by the qHGL approximation method. Although the generalization of the neutrino emissivity to non-plane-wave condensates in the presence of external magnetic fields is currently not available, the qHGL can be used to determine the favored phase for a given density and magnetic field strength, an essential input for future neutrino emissivity estimates from neutron star cores. Other physics details, such as the role of vector interactions, $\beta$-equilibrium and electrical neutrality, are left for future work.

\begin{acknowledgements}
F. Anzuini thanks Stefano Carignano for useful discussions and suggestions. This work is supported by The University of Melbourne with a Melbourne Research Scholarship and by funding from an Australian Research Council Discovery Project grant (DP170103625).
\end{acknowledgements}

\appendix

\section{Neutron star applications}
\label{sec:Applications}
\setcounter{equation}{0}
In this appendix, we discuss briefly extensions of our calculations to typical neutron star conditions. For simplicity, we focus on the CDW modulation. In \ref{sec:Asymmetric} we consider $\beta$-equilibrated quark matter. In \ref{sec:Chiral_limit} we discuss nonzero quark bare masses ($m_q \neq 0$). The latter affect the condensate parameters and reduce the size of density region where inhomogeneous phases are thermodynamically favored \cite{Maedan_2010, Buballa_2019_2}, i.e. the region where neutrino emission due to quark beta decay is allowed.

\subsection{Quark matter in $\beta$-equilibrium}
\label{sec:Asymmetric}
To apply our technique to neutron star matter, both $\beta$-equilibrium and electrical neutrality should be considered, strictly speaking \cite{Yakovlev_2001, Carignano_2015}. Additionally, one should consider a mixture of quarks and leptons \cite{Carignano_2015}. Broadly speaking, however, our conclusions do not change qualitatively, when these features are included. The case of asymmetric, $\beta$-equilibrated quark matter with electrons has been considered already in other works (see \cite{Buballa_2016, Carignano_2015} for example). Considering a mixture of up and down quarks as well as electrons, $\beta$-equilibrium amounts to having two different chemical potentials for up and down quarks ($\mu_u$ and $\mu_d$ respectively), which are related by

\begin{eqnarray}
\mu_u = \mu - \frac{2}{3}\mu_e, \qquad \mu_d = \mu + \frac{1}{3}\mu_e
\label{eq:chem_lepton}
\end{eqnarray}
where $\mu$ is the quark number chemical potential and $\mu_e$ denotes the electron chemical potential, determined by the condition of electrical neutrality. In our case, this amounts to rewriting the total free energy as

\begin{eqnarray}
\mathcal{F} = \mathcal{F}_e + \mathcal{F}_{\rm{qHGL}} \ ,
\label{eq:lepton}
\end{eqnarray}
where $\mathcal{F}_e$ denotes the electron free energy and the qHGL approximation reads
\begin{equation}
    \mathcal{F}_{\rm{qHGL}} = \mathcal{F}_{\rm{qHGL}}(\Delta_u, K_u, \mu_u, H) + \mathcal{F}_{\rm{qHGL}}(\Delta_d, K_d, \mu_d, H) \ ,
    \label{eq:qHGL_beta}
\end{equation}
where $\Delta_u, K_u$ denote the condensate parameters for up quarks, and $\Delta_d, K_d$ the corresponding ones for down quarks. To calculate the condensate parameters for the up and down quarks, one has to minimize Eqs. \eqref{eq:lepton} and \eqref{eq:qHGL_beta} with the condition in Eq. \eqref{eq:chem_lepton}, imposing that the system is electrically neutral.

\subsection{Expansion away from the chiral limit}
\label{sec:Chiral_limit}

Direct Urca neutrino emission due to quark beta decay is prohibited by kinematic arguments in free quark matter, but can activate in inhomogeneous quark matter. Nonzero bare quark masses reduce the size of the density region where quark matter is inhomogeneous, and hence reduce the stellar volume where direct Urca emission due to quark beta decay is possible. For nonzero bare quark masses, Eq. \eqref{eq:eff_mass} becomes 

\begin{eqnarray}
M(\bold{x}) = m_q - 2 G_S \Big[\braket{\bar \psi \psi}(\bold{x}) + \braket{\bar\psi i \gamma^5 \tau^3  \psi}(\bold{x})\Big] \ ,
\end{eqnarray}
where $m_q \neq 0$ denotes the bare quark mass. In the qHGL approximation, one needs to modify both the Taylor expansion to obtain the gradient terms and the homogeneous term in Eq. \eqref{eq:qHGL}. The homogeneous term in the qHGL approximation for nonzero $m_q$ can be written as \cite{Maedan_2010}

\begin{eqnarray}
    \mathcal{F}_{\rm{hom}}(M^{\prime}, m_q) =  \mathcal{F}^{\rm{vac}}_{\rm{hom}}(M^{\prime}) + \mathcal{F}^{\rm{med}}_{\rm{hom}}(M^{\prime}) + \frac{|M^{\prime} - m_q|^2}{4 G_S} \ , \nonumber \\
    &&
\end{eqnarray}
with $M^{\prime} = m_q + \Delta$. In the chiral limit, the Taylor expansion in Eq. \eqref{eqn:Kappaexp} is performed by calculating the derivatives of the free energy for $\Delta = 0$, i.e. close to the chiral restoration transition, with $M(\bold{x}) = 0$. On the other hand, as studied in \cite{Buballa_2019_2}, the Taylor expansion cannot be performed around $M(\bold{x}) = 0$ for $m_q \neq 0$, because chiral restoration is not reached due to the presence of the nonzero bare quark mass $m_q$. Accordingly, the authors in \cite{Buballa_2019_2} find (via standard Ginzburg-Landau techniques) that the free energy density contains odd powers of $M$ proportional to some coefficients $\gamma_{2j + 1}$ that do not appear for $m_q = 0$. In our case, if the Taylor expansion in Eq. \eqref{eqn:Kappaexp} is not centered at $M (\bold{x}) = 0$ the term $\partial\mathcal{F}/\partial(\Delta^2)$ cannot be calculated with the technique developed in this work, and the coefficients $\gamma_{2j + 1}$ proportional to odd powers of the condensate parameters cannot be obtained.

In summary, the extension of our results to the case of finite masses is interesting but nontrivial, and will be addressed in future work. 

\section{Expansion for small magnetic fields}
\label{sec:H_expansion}
\setcounter{equation}{0}
The free energy of the higher Landau levels (Eq. \eqref{eq:F_HLL}) is a function of $|e_f H|$. The expansion of Eq. \eqref{eq:F_2} for small magnetic fields requires caution, since the absolute value is not differentiable at $e_f H = 0$. Hence, a mathematical prescription is necessary to ensure the validity of the Taylor expansion for small magnetic fields.

We replace the absolute value $|x|$ function with
\begin{eqnarray}
|x| \mapsto f(x, k) = \frac{2}{k}\ln(1 + e^{k x}) - x - \frac{2}{k}\log(2)
\label{eq:module}
\end{eqnarray}
where the $k$ parameter determines the steepness of the curve and the constant term ensures that $f(0, k) = 0$. The derivative is given by the logistic sigmoid function
\begin{eqnarray}
\frac{d|x|}{dx} \mapsto  f^{\prime}(x, k) =  \frac{2}{1 + e^{-k x}} - 1
\label{eq:module_derivative}
\end{eqnarray}
which is continuous in $x = 0$. 

\begin{figure}
  \includegraphics[width=8.5 cm, height = 13 cm]{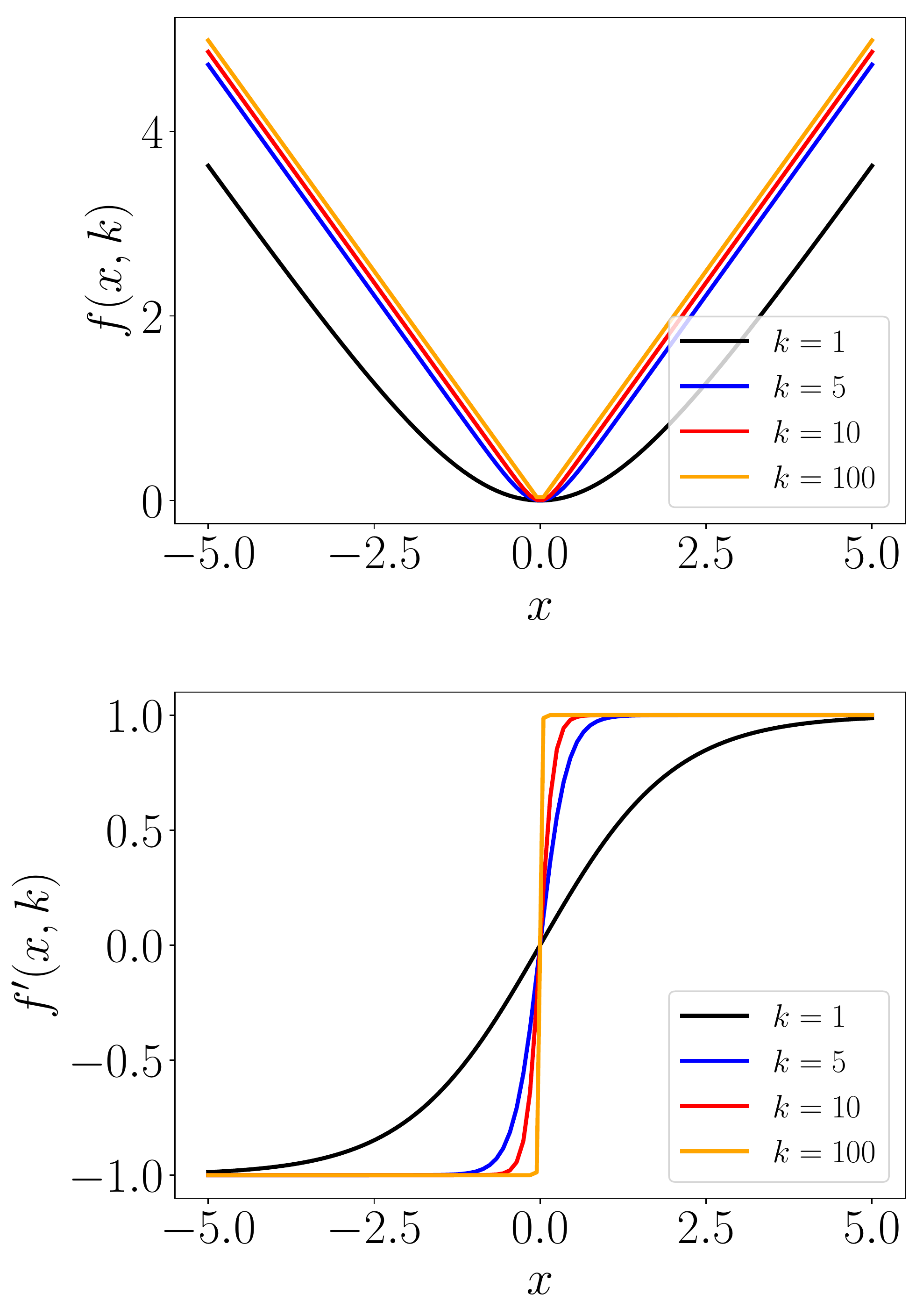}
\caption{Smoothing functions Eqs. (\ref{eq:module}) and (\ref{eq:module_derivative}) versus $x$ for different various values of the steepness parameter $k$.}
\label{fig:Sigmoid}       
\end{figure}

Figure \ref{fig:Sigmoid} shows the plots of $f(x, k)$ and $f^{\prime}(x, k)$ as functions of $x$ for different values of $k$. For $k \approx 100$ both $f$ and $f^{\prime}$ reproduce well the absolute value and the sign function of $x$ respectively, with $f$ differentiable in the origin. Hence, when expanding the quark free energy for small magnetic fields, one has to replace $|e_f H|$ with $f(e_f H, k)$, so that $f(x, k)$ and $f^{\prime}(x, k)$ reproduce the absolute value of the magnetic field and its derivative, and the Taylor expansion centered in $e_f H = 0$ is mathematically consistent. Additionally, we notice that when performing the Taylor expansion of Eq.(\ref{eq:F_2}) (and similarly for the vacuum term), the terms proportional to odd powers of the magnetic field vanish, since they are proportional to $f^{\prime}(0, k) = 0$. This ensures that the qHGL functional is independent of the magnetic field sign and therefore preserves the rotational symmetry of the system \cite{Frolov_2010, Abuki_2018}.

\section{Coefficients of the qHGL expansion}
\label{sec:Coefficients}
\setcounter{equation}{0}
We report here the coefficients $\gamma_j$ introduced in Eq. (\ref{eq:qHGL}). Since $\gamma_j$ is obtained by expanding the free energy for small magnetic fields, it is given by the sum of the magnetic field-independent contribution plus corrections proportional to even powers of the magnetic field \cite{Abuki_2018}, as discussed in \ref{sec:H_expansion}. For $H = 0$, the formulas below for $\gamma_j$ reproduce the coefficients found in \cite{Carignano_2018}. Including only corrections up to $(e_fH)^2$, the coefficients are given by
\begin{eqnarray}
\gamma_4(\mu,\Lambda, H)   = && -\frac{N_c}{16\pi^2}\sum_{f}\bigg[\log\Big(\frac{32\mu^2}{3\Lambda^2}\Big)  \nonumber \\
&& + (2 e_f H)^2\Big( \frac{85}{72 \Lambda^4} + \frac{1}{16 \mu^4}\Big)\bigg] 
\\
\nonumber\\
\gamma_6(\mu,\Lambda, H)  = && \frac{N_c}{32\pi^2}\sum_{f}\bigg[\Big(\frac{11}{9\Lambda^2}+\frac{1}{3\mu^2}\Big) \nonumber \\
&& + (2 e_f H)^2\Big( \frac{575}{324 \Lambda^6} - \frac{1}{24 \mu^6}\Big)\bigg]
\\
 \nonumber\\
\gamma_8(\mu,\Lambda, H)  = && \frac{N_c}{256\pi^2}\sum_{f}\bigg[\Big(\frac{1}{2\mu^4} - \frac{85}{27\Lambda^4}\Big) \nonumber \\
&& - (2 e_f H)^2\Big( \frac{18305}{1620 \Lambda^8} + \frac{5}{32 \mu^8}\Big)\bigg]
\\
 \nonumber\\
\gamma_{10}(\mu,\Lambda, H)  = && \frac{N_c}{1024\pi^2}\sum_{f}\bigg[\Big(\frac{230}{567\Lambda^6}+\frac{1}{21\mu^6}\Big) \nonumber \\
&& + (2 e_f H)^2\Big( \frac{3233}{1215 \Lambda^{10}} - \frac{1}{40 \mu^{10}}\Big)\bigg]
\end{eqnarray}

\begin{eqnarray}
\gamma_{12}(\mu,\Lambda, H)  = && \frac{N_c}{4096\pi^2}\sum_{f}\bigg[\Big(\frac{1}{36\mu^8} - \frac{1046}{3645\Lambda^8}\Big) \nonumber \\
&&   - (2 e_f H)^2\Big( \frac{137845}{45927 \Lambda^{12}} + \frac{1}{48 \mu^{12}}\Big)\bigg] \  \ 
\\
 \nonumber\\
\gamma_{14}(\mu,\Lambda, H)  = && \frac{N_c}{16384\pi^2}\sum_{f}\bigg[\Big(\frac{1}{55\mu^{10}} + \frac{25864}{120285 \Lambda^{10}}\Big) \nonumber \\
&&   + (2 e_f H)^2\Big(\frac{1666750}{505197 \Lambda^{14}} - \frac{1}{56 \mu^{14}}\Big)\bigg] \ , \nonumber \\  
\\
\nonumber
\gamma_{16}(\mu,\Lambda, H)  = && \frac{N_c}{65536\pi^2}\sum_{f}\bigg[\Big(\frac{1}{78\mu^{12}} - \frac{1102760}{6567561 \Lambda^{12}}\Big) \nonumber \\
&&   - (2 e_f H)^2\Big(\frac{912622}{255879 \Lambda^{16}} + \frac{1}{64 \mu^{16}}\Big)\bigg] \ , \nonumber \\  
\end{eqnarray}
where the sum over the index $f$ runs over the quark flavor.

\section{Numerical accuracy}
\label{sec:Comparison}

\begin{figure*}
  \includegraphics[width=17.5 cm, height = 6.3 cm]{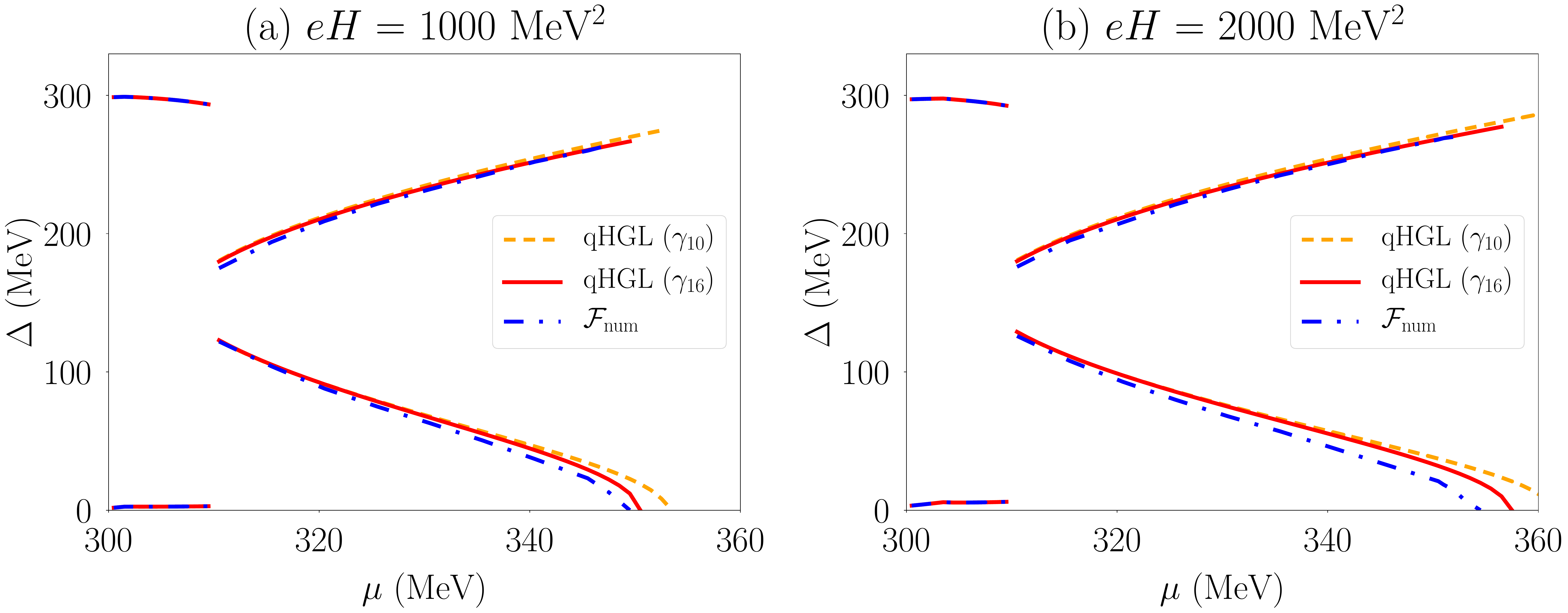}
\caption{Accuracy of the qHGL approximation: comparison of CDW $\Delta$ and $K$  obtained with the qHGL approximation and $\mathcal{F}_{\textrm{num}}$ for two typical magnetic field strengths ($eH = 1000$ MeV$^2$ and $eH = 2000$ MeV$^2$). \textit{(a)} $eH = 1000$ MeV$^2$. \textit{(b)} $eH = 2000$ MeV$^2$. In both panels, we include corrections in the qHGL functional up to $\gamma_{10}$ (orange, dashed lines) or $\gamma_{16}$ (red solid lines).}
\label{fig:numerical}       
\end{figure*}
We can compare the qHGL approximation against \enquote{more exact} quantities. Our work involves two approximations for the free energy: \textit{(i)} Taylor expansion in the condensate parameters; and \textit{(ii)} continuum approximation for the discrete energy levels. For the latter, we follow the literature and expand also for small magnetic fields (cf. \cite{Abuki_2018} for example). Let us denote with $\mathcal{F_{\rm{num}}}$ the free energy obtained applying only the approximation in point \textit{(ii)}. Let us denote with $\mathcal{F^{\rm{PT}}_{\rm{exact}}}$ the exact free energy density regularized with the proper-time regularization scheme, calculated with neither the approximation in point \textit{(i)} nor \textit{(ii)} \cite{Frolov_2010, Carignano_2015, Nishiyama_2015}. We test quantitatively the accuracy of the qHGL approximation by comparing the condensate parameters obtained with the qHGL expansion with $\mathcal{F_{\rm{num}}}$, and qualitatively with $\mathcal{F^{\rm{PT}}_{\rm{exact}}}$.

In Figures \ref{fig:numerical}(a) and \ref{fig:numerical}(b) we compare the amplitude $\Delta$ and parameter $K$ of the CDW modulation obtained with the qHGL functional including gradient terms up to $\gamma_{10}$ or $\gamma_{16}$ (solid red and dashed orange lines respectively) and with $\mathcal{F}_{\rm{num}}$ (blue, dashed-dotted line) calculated for $eH = 1000$ MeV$^2$ and $eH = 2000$ MeV$^2$. The qHGL approximation results show a good agreement with $\mathcal{F}_{\rm{num}}$ (either including terms up to $\gamma_{10}$ or $\gamma_{16}$) for $300 \ \textrm{MeV} \lesssim \mu \lesssim 340$ MeV. For $\mu \lesssim 310.5$ MeV, the blue, orange and red curves are degenerate. Close to the first-order phase transition, the condensate parameters are determined to within $\lesssim 3$ per cent. For $eH = 1000$ MeV$^2$ and $\mu \lesssim 340$ MeV we find that the typical relative error in $\Delta$ is within $\lesssim 10$ per cent and in $K$ to within $\lesssim 1$ per cent. For $\mu \lesssim 340$ MeV and stronger fields ($eH = 2000$ MeV$^2$) the relative error in $\Delta$ is within $\lesssim 13$ per cent and in $K$ to within $\lesssim 2$ per cent. An improved version of the qHGL expansion will be presented in future work.

It is interesting to compare the results obtained with the Pauli-Villars and proper-time regularization schemes, as discussed briefly in Sections \ref{sec:Benchmark_Mag} and \ref{sec:Solitons_Mag}. The two schemes correspond to related but different physics \citep{Klevansky_1992}, so one should not expect detailed quantitative agreement. However, one does hope for broad qualitative agreement, to confirm that the main conclusions depend weakly on regularization, and indeed this is what we find. We compare the results obtained with the qHGL approximation with $\mathcal{F^{\rm{PT}}_{\rm{exact}}}$ calculated in the literature, where the case of ultra-strong magnetic fields is usually considered \cite{Frolov_2010, Nishiyama_2015, Carignano_2015, Buballa_2016}. For the CDW, the typical values of $\mu_c, \Delta$ and $K$ are similar for $eH = 0$. For example, using the parameters adopted in \cite{Nishiyama_2015}, one finds that the effective quark mass in vacuum is $\approx 330$ MeV, and $\mu_c \approx 350$ MeV, in contrast to our case (with $M \approx 300$ MeV and $\mu_c \approx 345$ MeV, \cite{Carignano_2018}). However qualitative features in $\mathcal{F^{\rm{PT}}_{\rm{exact}}}$ are preserved by the qHGL approximation for $eH > 2000$ MeV$^2$ and $T = 0$ , such as the suppression of chiral restoration. For strong magnetic fields, one can neglect the magnetic corrections of the higher Landau levels in the qHGL approximation and include magnetic effects only via the LLL (i.e. the leading-order contribution). For example, we find that for $eH \gtrsim 8000$ MeV$^2$ (not shown here), the CDW parameters obtained with the qHGL approximation are in qualitatively agreement with the parameters obtained with  $\mathcal{F^{\rm{PT}}_{\rm{exact}}}$ \cite{Frolov_2010}, showing that chiral restoration is not reached up to $\mu \approx 400$ MeV.

\bibliographystyle{spphys}       
\bibliography{BIBLIO}   

\end{document}